\documentclass[conference]{IEEEtran}
\IEEEoverridecommandlockouts
\usepackage{cite}
\usepackage{amsmath,amssymb,amsfonts}
\usepackage{graphicx}
\usepackage{textcomp}
\usepackage{xcolor}
\usepackage{float}

\begin{document}

\title{Scaling Legal AI: Benchmarking State-Space Models and
Transformers for Statutory Classification and Case Law Retrieval\\
}

\author{\IEEEauthorblockN{Anuraj Maurya}
\IEEEauthorblockA{Bengaluru, India\\
anurajmaurya@iisc.ac.in}}

\maketitle

\begin{abstract}
Statutory corpora and judicial decisions are growing faster than legal
professionals can read them, and the documents involved are long: individual
judgments routinely exceed the context window of standard encoder models.
Transformer-based architectures dominate current legal NLP benchmarks, but
their quadratic attention cost forces practitioners to truncate or fragment
exactly the documents that most require whole-document reasoning. Selective
state-space models (SSMs) such as Mamba offer linear-time sequence modelling
and are therefore a natural candidate for this setting, yet they have not been
systematically evaluated on legal classification and retrieval.

We present a preliminary benchmark of two SSM variants (Mamba and SSD-Mamba)
against three transformer encoders (BERT, DeBERTa, and Longformer) on four
legal classification tasks (ECtHR, EUR-Lex, SCOTUS, ILDC/ILC) and two
case-retrieval tasks (ECtHR, ILDC), using a shared windowing and aggregation
pipeline. In these initial results the strongest SSM is within roughly 1.3
points of the strongest transformer on every task and metric, and SSD-Mamba
attains the best score on most metrics for ECtHR classification, ILDC
classification, and ECtHR retrieval, while processing approximately 3$\times$
more tokens per second than DeBERTa and 4$\times$ more than Longformer.
DeBERTa remains strongest on SCOTUS and on EUR-Lex F1. Because results are
reported without variance estimates across random seeds or significance
testing, we present them as preliminary evidence and as a concrete analysis
agenda rather than a definitive ranking, and we describe the controls planned
for a subsequent version.
\end{abstract}


\section{Introduction}

Legal practice is a reading problem at a scale that no longer fits human
capacity. National and supranational databases hold millions of statutes and
decisions, and the annual increment alone exceeds what any research team can
review. The documents themselves compound the difficulty. A single appellate
judgment may run to hundreds of pages, and the reasoning that determines its
outcome is frequently distributed across widely separated passages, with facts
recited early and their legal significance resolved much later. Statutory
language adds open-texture terms, dense cross-referencing, and layered
exceptions that defeat purely lexical retrieval~\cite{ashley2017analytics,
benchcapon2022winter}.

Modern legal NLP addresses this with transformer
encoders~\cite{vaswani2017attention}, which have produced substantial gains on
statutory tagging, judgment prediction, and semantic
retrieval~\cite{chalkidis2021lexglue, chalkidis2020legalbert, malik2021ildc,
aletras2016ecthr, katz2017supreme}. The architecture, however, carries a
structural mismatch with the domain. Attention cost grows quadratically in
sequence length, so long judgments must be truncated, or split into windows
that are encoded independently and then pooled. Sparse-attention variants such
as Longformer~\cite{beltagy2020longformer} and BigBird~\cite{zaheer2020bigbird}
raise the ceiling to a few thousand tokens without removing it, and
hierarchical encoders recover much of the lost signal at the cost of a second
modelling stage~\cite{chalkidis2021lexglue}. In every case the model sees the
document in pieces.

Selective state-space models change the shape of this trade-off. Rather than
computing pairwise token interactions, an SSM propagates a hidden state through
the sequence with input-dependent gating, giving linear time and constant
memory in sequence length~\cite{gu2021lssl, gu2022s4, gu2023mamba}. Mamba and
its structured state-space duality successor
(Mamba-2/SSD)~\cite{gu2023mamba, dao2024ssd} reach transformer-competitive
quality on language modelling while admitting far longer inputs, and have begun
to appear in retrieval settings~\cite{zhang2024mambaretriever}. Whether that
advantage survives contact with legal text is an open question, and it is not
obvious that it should: SSMs are measurably weaker than attention at exact
copying and associative recall~\cite{jelassi2024copying}, both of which
plausibly matter for citation-heavy statutory reasoning.

\subsection{Research Questions}
This preprint addresses the first of three questions directly and frames the
remaining two as hypotheses for subsequent work.

\begin{itemize}
  \item \textbf{RQ1 (Accuracy).} Under a shared preprocessing and aggregation
  pipeline, do selective SSMs match transformer encoders on statutory
  classification and case-law retrieval?
  \item \textbf{RQ2 (Scaling).} How do accuracy and throughput behave as a
  function of document length, and does reduced window fragmentation account
  for any SSM advantage?
  \item \textbf{RQ3 (Failure modes).} Do the two paradigms fail differently,
  for example by label frequency or document length, and what does that imply
  for deployment?
\end{itemize}

\subsection{Contributions}
\begin{itemize}
  \item \textbf{An architectural comparison on legal tasks.} Five neural
  systems across six legal tasks, with windowing overlap, pooling, and
  evaluation held constant across architectures.
  \item \textbf{Accuracy--throughput measurements.} Task-level accuracy and
  ranking metrics reported alongside throughput, showing that the strongest SSM
  and the strongest transformer are separated by small accuracy margins but a
  large throughput gap.
  \item \textbf{A deployment-oriented interpretation.} A simple cost model
  translating throughput into corpus-indexing time.
  \item \textbf{A falsifiable analysis agenda.} Explicit hypotheses about
  document length, window fragmentation, and rare-label performance, together
  with the experiments that would confirm or refute them.
\end{itemize}

SSMs have been benchmarked for general long-document
classification~\cite{ssmlongdoc2024} and dense
retrieval~\cite{zhang2024mambaretriever}. To our knowledge this is the first
comparison focused on \emph{statutory classification and case-law retrieval}
specifically.

\section{Related Work}

\subsection{Legal NLP Benchmarks and Domain Models}
LexGLUE~\cite{chalkidis2021lexglue} consolidated seven English legal datasets,
including ECtHR violation prediction, SCOTUS issue-area classification, and
EUR-Lex EuroVoc tagging, into a standard suite. Two of its findings bear on any
architectural comparison. First, legal-domain pretraining helps:
LEGAL-BERT~\cite{chalkidis2020legalbert} and
CaseLaw-BERT~\cite{zheng2021casehold} outperform generic encoders,
particularly on macro-averaged metrics that weight infrequent labels. Second,
long-document handling matters independently of pretraining: hierarchical
variants that encode segments and aggregate them substantially outperform
truncated encoders on the long-document tasks. The present comparison uses
generic-domain encoders only; adding legal-domain and hierarchical baselines is
the first priority for the next version (Section~\ref{sec:future}).

For Indian law, ILDC~\cite{malik2021ildc} provides judgment-outcome prediction
and explanation over Supreme Court judgments whose length considerably exceeds
ECtHR facts sections. Recent hierarchical and explainability-oriented
architectures report strong absolute ECtHR results~\cite{lexfaith2025}; we
position this work as an architectural comparison under a fixed pipeline rather
than a state-of-the-art claim on any individual task. Generative legal-domain
models~\cite{colombo2024saullm} and broad legal-reasoning
benchmarks~\cite{guha2023legalbench} have shifted much recent attention toward
decoder LLMs. Efficient encoders nonetheless remain the practical choice for
corpus-scale classification and first-stage retrieval, which is the regime we
study.

\subsection{Long-Context Transformers}
Sparse and structured attention reduce the quadratic term without eliminating
it. Longformer~\cite{beltagy2020longformer} combines sliding-window with global
attention; BigBird~\cite{zaheer2020bigbird} adds random attention with
theoretical coverage guarantees; earlier sparse
factorisations~\cite{child2019sparse} established the general strategy. In
legal NLP the dominant practical alternative is hierarchical encoding, in which
segment representations from a standard encoder are aggregated by a
second-stage model~\cite{chalkidis2021lexglue}. Long-context embedding
models~\cite{gunther2023jina} extend the same idea to retrieval. The sliding
window with overlap and mean pooling used here is a deliberately simple member
of this family, chosen because it can be applied identically to every
architecture under test.

\subsection{State-Space and Hybrid Architectures}
Linear state-space layers~\cite{gu2021lssl} and S4~\cite{gu2022s4} established
that structured recurrences can model long-range dependencies at subquadratic
cost. Mamba~\cite{gu2023mamba} added input-dependent selectivity and a
hardware-aware parallel scan; Mamba-2~\cite{dao2024ssd} reframed the mechanism
through structured state-space duality, yielding the SSD formulation we
evaluate as SSD-Mamba. Applications now span long-document
classification~\cite{ssmlongdoc2024} and dense
retrieval~\cite{zhang2024mambaretriever}.

The limitations are equally documented. Transformers outperform SSMs on copying
and associative recall~\cite{jelassi2024copying}. Partly in response, the field
has converged on hybrids that interleave attention with SSM layers, including
Jamba and Jamba-1.5~\cite{lieber2024jamba, jambateam2024jamba15},
Samba~\cite{ren2024samba}, Zamba~\cite{glorioso2024zamba}, and parallel-head
variants such as Hymba~\cite{dong2024hymba}. We return to the relevance of
these designs in Section~\ref{sec:analysis}.

\section{Methodology}

\subsection{Shared Pipeline}
\begin{figure}[t]
\centering
\includegraphics[width=\columnwidth]{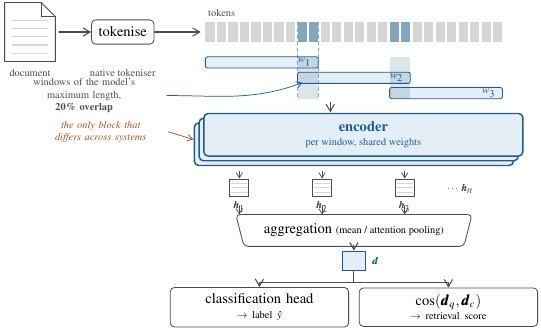}
\vspace{1.0em}
\caption{The shared processing pipeline. Only the encoder block differs
across the systems under comparison.}
\label{fig:pipeline}
\end{figure}

All systems share one pipeline so that differences are attributable as far as
possible to architecture rather than preprocessing. A document is tokenised
with the tokeniser native to its model, segmented into windows of the model's
supported length with 20\% overlap between consecutive windows, encoded
window-by-window, and reduced to a single document representation by
aggregation. Classification heads are applied to the aggregated
representation; retrieval scores are cosine similarities between aggregated
query and candidate representations. Fig.~\ref{fig:pipeline} summarises the
flow.

\subsection{Design Choices}
\textbf{Overlap.} Independent windows sever cross-boundary dependencies, and a
decisive clause landing on a boundary is otherwise split across two
representations. A 20\% overlap guarantees that any span shorter than one fifth
of the window length appears intact in at least one window, at the cost of
approximately 25\% more encoder calls.

\textbf{Aggregation.} For multi-label classification, per-window sigmoid
probabilities are averaged before thresholding; for single-label
classification, logits are averaged before softmax; for retrieval, window
embeddings are mean-pooled. Mean pooling treats every window as equally
informative, which is a conservative assumption for legal documents whose
decisive passages are localised. Max pooling and attention pooling are natural
alternatives that we have not yet evaluated.

\subsection{Context Length}
BERT and DeBERTa admit 512 tokens and Longformer admits 4{,}096. Mamba and
SSD-Mamba have no architectural maximum context length and are therefore
applied with longer windows than the transformer baselines, which reduces the
number of independently encoded fragments per document.

\subsection{Tasks and Data}
Table~\ref{tab:datasets} summarises the six tasks. ECtHR requires multi-label
prediction of allegedly violated Convention articles from a judgment's facts
section. EUR-Lex requires multi-label EuroVoc tagging of EU legislation.
SCOTUS requires single-label issue-area classification. ILDC/ILC covers
statute tagging and outcome prediction over Indian Supreme and High Court
judgments, and is our longest-input condition. The two retrieval tasks require
ranking candidate precedents given a query judgment.

Split construction differs across datasets: official LexGLUE splits are used
where available, and author-constructed splits otherwise. Absolute numbers are
therefore intended for comparison \emph{across architectures within this
study}, and should not be compared directly with published leaderboard results.

\begin{table}[t]
\centering
\caption{Task summary.}
\label{tab:datasets}
\footnotesize
\begin{tabular}{lll}
\hline
\textbf{Dataset} & \textbf{Task} & \textbf{Metrics} \\
\hline
ECtHR      & Multi-label cls.  & Micro/Macro-F1, Acc., AUC \\
EUR-Lex    & Multi-label cls.  & Micro/Macro-F1, Acc., AUC \\
SCOTUS     & Single-label cls. & Micro/Macro-F1, Acc., AUC \\
ILDC/ILC   & Multi-task cls.   & Micro/Macro-F1, Acc., AUC \\
ECtHR-LCR  & Retrieval         & MAP, MRR, R@10, nDCG@10 \\
ILDC-Ret.  & Retrieval         & MAP, MRR, R@10, nDCG@10 \\
\hline
\end{tabular}
\end{table}

\begin{table}[t]
\centering
\caption{Systems under comparison.}
\label{tab:models}
\footnotesize
\begin{tabular}{llc}
\hline
\textbf{System} & \textbf{Family} & \textbf{Max ctx.} \\
\hline
BERT        & Transformer (full attention)   & 512 \\
DeBERTa     & Transformer (full attention)   & 512 \\
Longformer  & Transformer (sparse attention) & 4{,}096 \\
Mamba       & Selective SSM                  & none \\
SSD-Mamba   & Selective SSM (Mamba-2/SSD)    & none \\
\hline
\end{tabular}
\end{table}

\section{Results}

Tables~\ref{tab:ecthr}--\ref{tab:ildc_retrieval} report results for all
systems. The best value in each column is shown in bold. Values are point
estimates; variance across random seeds and statistical significance have not
yet been assessed, and differences below roughly one point should be read
accordingly. Throughput (Tok/s) was measured for inference on a single
accelerator and is intended for relative comparison between systems.

\begin{table*}[!t]
\centering
\begin{minipage}{0.48\textwidth}
\centering
\caption{ECtHR multi-label classification.}
\label{tab:ecthr}
\scriptsize
\begin{tabular}{lccccc}
\hline
\textbf{Model} & \textbf{Micro-F1} & \textbf{Macro-F1} & \textbf{Acc.} & \textbf{AUC} & \textbf{Tok/s} \\
\hline
BERT        & 72.3 & 65.1 & 68.4 & 75.2 & 18k \\
Longformer  & \textbf{76.1} & 70.4 & 71.9 & 78.6 & 12k \\
DeBERTa     & 75.8 & 69.9 & 72.1 & 79.0 & 16k \\
Mamba       & 75.2 & 70.1 & 71.7 & 78.4 & 39k \\
SSD-Mamba   & 76.0 & \textbf{71.3} & \textbf{73.0} & \textbf{80.1} & \textbf{46k} \\
\hline
\end{tabular}
\end{minipage}
\hfill
\begin{minipage}{0.48\textwidth}
\centering
\caption{EUR-Lex multi-label classification.}
\label{tab:eurlex}
\scriptsize
\begin{tabular}{lccccc}
\hline
\textbf{Model} & \textbf{Micro-F1} & \textbf{Macro-F1} & \textbf{Acc.} & \textbf{AUC} & \textbf{Tok/s} \\
\hline
BERT        & 69.5 & 60.8 & 65.2 & 72.0 & 17k \\
Longformer  & 73.5 & 65.4 & 68.1 & 75.5 & 11k \\
DeBERTa     & \textbf{74.2} & \textbf{66.7} & 68.9 & \textbf{76.4} & 15k \\
Mamba       & 72.8 & 65.9 & 68.2 & 75.9 & 36k \\
SSD-Mamba   & 73.9 & 66.4 & \textbf{69.3} & 76.1 & \textbf{44k} \\
\hline
\end{tabular}
\end{minipage}
\end{table*}

\begin{table*}[!t]
\centering
\begin{minipage}{0.48\textwidth}
\centering
\caption{SCOTUS issue-area classification.}
\label{tab:scotus}
\scriptsize
\begin{tabular}{lccccc}
\hline
\textbf{Model} & \textbf{Micro-F1} & \textbf{Macro-F1} & \textbf{Acc.} & \textbf{AUC} & \textbf{Tok/s} \\
\hline
BERT        & 79.6 & 79.4 & 80.1 & 83.1 & 19k \\
Longformer  & 81.8 & 81.0 & 82.4 & 85.2 & 13k \\
DeBERTa     & \textbf{83.8} & \textbf{82.7} & \textbf{84.0} & \textbf{86.7} & 17k \\
Mamba       & 82.5 & 81.5 & 83.1 & 86.0 & 42k \\
SSD-Mamba   & 83.1 & 82.0 & 83.7 & 86.3 & \textbf{49k} \\
\hline
\end{tabular}
\end{minipage}
\hfill
\begin{minipage}{0.48\textwidth}
\centering
\caption{ILDC/ILC statute tagging and outcome prediction.}
\label{tab:ildc}
\scriptsize
\begin{tabular}{lccccc}
\hline
\textbf{Model} & \textbf{Micro-F1} & \textbf{Macro-F1} & \textbf{Acc.} & \textbf{AUC} & \textbf{Tok/s} \\
\hline
BERT        & 68.9 & 61.3 & 73.4 & 76.5 & 15k \\
Longformer  & 72.7 & 65.5 & 75.9 & 79.0 & 11k \\
DeBERTa     & 73.0 & 66.2 & 76.1 & 79.7 & 14k \\
Mamba       & 72.8 & 65.8 & 76.8 & 80.2 & 34k \\
SSD-Mamba   & \textbf{73.5} & \textbf{66.5} & \textbf{77.1} & \textbf{80.8} & \textbf{41k} \\
\hline
\end{tabular}
\end{minipage}
\end{table*}

\begin{table*}[!t]
\centering
\begin{minipage}{0.48\textwidth}
\centering
\caption{Legal case retrieval, ECtHR.}
\label{tab:lcr}
\scriptsize
\begin{tabular}{lccccc}
\hline
\textbf{Model} & \textbf{MAP} & \textbf{MRR} & \textbf{R@10} & \textbf{nDCG@10} & \textbf{Tok/s} \\
\hline
BERT        & 64.1 & 65.3 & 68.9 & 66.2 & 14k \\
Longformer  & 67.8 & 68.6 & \textbf{72.5} & 69.5 & 10k \\
DeBERTa     & 67.2 & 68.0 & 71.8 & 69.0 & 13k \\
Mamba       & 68.5 & 69.3 & 72.1 & 70.4 & 35k \\
SSD-Mamba   & \textbf{69.1} & \textbf{70.2} & 72.3 & \textbf{71.0} & \textbf{41k} \\
\hline
\end{tabular}
\end{minipage}
\hfill
\begin{minipage}{0.48\textwidth}
\centering
\caption{Case retrieval, ILDC (India).}
\label{tab:ildc_retrieval}
\scriptsize
\begin{tabular}{lccccc}
\hline
\textbf{Model} & \textbf{MAP} & \textbf{MRR} & \textbf{R@10} & \textbf{nDCG@10} & \textbf{Tok/s} \\
\hline
BERT        & 60.5 & 61.8 & 65.1 & 63.2 & 13k \\
Longformer  & \textbf{65.2} & \textbf{66.1} & 69.5 & 67.3 & 9k \\
DeBERTa     & 64.8 & 65.5 & \textbf{70.1} & 67.8 & 12k \\
Mamba       & 64.5 & 65.0 & 69.7 & 67.0 & 31k \\
SSD-Mamba   & 65.0 & 65.8 & 70.0 & \textbf{68.1} & \textbf{38k} \\
\hline
\end{tabular}
\end{minipage}
\end{table*}

\noindent\textbf{Abbreviations.}
\begin{table}[H]
\centering
\renewcommand{\arraystretch}{1.1}
\footnotesize
\begin{tabular}{p{2.4cm}p{5.2cm}}
\textbf{Micro/Macro-F1} & Micro- and macro-averaged F1 \\
\textbf{Acc.}     & Classification accuracy \\
\textbf{AUC}      & Area under the ROC curve \\
\textbf{MAP}      & Mean average precision \\
\textbf{MRR}      & Mean reciprocal rank \\
\textbf{R@10}     & Recall at top-10 retrieved documents \\
\textbf{nDCG@10}  & Normalised discounted cumulative gain at 10 \\
\textbf{Tok/s}    & Throughput in tokens per second \\
\end{tabular}
\end{table}

\section{Analysis}
\label{sec:analysis}

\subsection{Accuracy Margins Are Small}
Across all six tasks and all reported metrics, the gap between the best
transformer and the best SSM never exceeds 1.3 points. SSD-Mamba leads on
three of four metrics for ECtHR classification, all four for ILDC/ILC, and
three of four for ECtHR retrieval. DeBERTa leads on all four SCOTUS metrics
and on EUR-Lex F1 and AUC. On ILDC retrieval the leading system differs by
metric, with Longformer ahead on MAP and MRR and SSD-Mamba ahead on nDCG@10.
Margins of this size are within the range that seed variance commonly produces
in fine-tuned legal classifiers, so the defensible reading at this stage is
that SSMs are \emph{competitive with} rather than \emph{superior to}
generic-domain transformer encoders on these tasks.

\subsection{Throughput Differences Are Large}
In contrast to the accuracy margins, throughput differences are substantial
and consistent. SSD-Mamba processes 2.9--3.2$\times$ as many tokens per second
as DeBERTa and 3.7--4.2$\times$ as many as Longformer across the six tasks.
Unlike the accuracy differences, these ratios are large enough that seed
variance is unlikely to change their direction. The combination of near-parity
accuracy and a 3--4$\times$ throughput gap is the central empirical
observation of this preprint.

\subsection{SSD-Mamba Versus Mamba}
SSD-Mamba matches or exceeds Mamba on every metric of every task, while also
achieving higher throughput. Within the SSM family, the Mamba-2/SSD
formulation therefore appears to be the stronger default for this setting.

\subsection{Where Transformers Lead}
DeBERTa's clearest advantage is on SCOTUS, where it leads on every metric.
One possible explanation is that the benefit of long-context processing is
smallest on tasks whose decisive signal is localised, so that the stronger
pretraining of DeBERTa dominates. A second is the documented weakness of SSMs
at exact recall of sparse, specific associations~\cite{jelassi2024copying}.
The present data cannot distinguish these explanations. We note, however, that
the macro-F1 results do not show a uniform transformer advantage on rare labels:
SSD-Mamba has the highest macro-F1 on ECtHR and ILDC/ILC, while DeBERTa leads
on EUR-Lex and SCOTUS.

\subsection{What These Results Do Not Yet Establish}
Three hypotheses motivate this work but are not tested here. First, that
\emph{document length} is the variable that discriminates between the
paradigms, which would predict that transformer performance falls across
length buckets while SSM performance remains comparatively flat. Second, that
reduced \emph{window fragmentation} is the mechanism behind any SSM advantage,
which requires measuring windows per document directly. Third, that attention
and state-space recurrence are \emph{complementary}, with attention favoured
where the decisive signal is local and specific and recurrence favoured where
evidence must be integrated across long spans. If that reading holds, it would
offer a domain-level explanation for why production systems increasingly
interleave the two mechanisms~\cite{lieber2024jamba, jambateam2024jamba15,
ren2024samba}. Each hypothesis is directly testable with the experiments listed
in Section~\ref{sec:future}.

\section{Deployment Considerations}

Efficiency claims become actionable only when expressed in the units that
institutions budget in. As an illustrative extrapolation from the throughput
figures above, consider a one-time indexing pass over a corpus of $10^{6}$
judgments averaging $10^{4}$ tokens, that is $10^{10}$ tokens, on a single
accelerator of the class used in our measurements. At Longformer's 10k
tokens/s this requires roughly $10^{6}$ seconds, about 11.6 days. At
SSD-Mamba's 41k tokens/s it requires roughly $2.4 \times 10^{5}$ seconds,
about 2.8 days. Re-indexing after a statutory amendment or a batch of new
decisions would therefore shift from a task scheduled around a fortnight to one
that fits a weekend. These figures assume throughput scales linearly to the
full corpus and exclude I/O and preprocessing overhead.

For a court registry, a legal-aid organisation, or a national statutory
database, where access to legal information is most constrained by budget, a
difference of this size can determine whether an index is maintained or built
once. Conversely, for precision-critical work on tasks resembling SCOTUS, our
results indicate that a well-chosen transformer remains the better default.

\subsection{Toward Explainable Predictions}
The aggregation scheme used here admits a natural, if partial, form of
attribution. Because a document-level prediction is the mean of window-level
predictions, the contribution of each window is directly recoverable, and
windows can be ranked by their influence on a predicted label. This permits
passage-level highlighting over the original judgment without additional
machinery. Window-level attribution localises evidence but does not explain the
reasoning that connects it, and should not be presented to practitioners as a
justification for an outcome.

\section{Limitations}

\textbf{Statistical robustness.} Results are point estimates without
variance across seeds or significance tests. Given the small accuracy margins,
the ranking of systems on individual tasks may change under repeated runs.

\textbf{Baseline coverage.} The comparison omits a legal-domain pretrained
encoder (e.g., LEGAL-BERT), a hierarchical transformer, and a lexical retrieval
baseline such as BM25~\cite{robertson2009bm25}. All three are standard in legal
NLP, and their absence means the transformer side of the comparison is not yet
at full strength.

\textbf{Parameter budgets.} Model sizes were not explicitly matched across
families, so some differences may reflect capacity rather than architecture.

\textbf{Splits.} Mixing official and author-constructed splits limits
comparability with published results.

\textbf{Jurisdictional and linguistic scope.} Our corpora are English-language
European, US, and Indian sources. We make no claim about civil-law systems,
non-English statutory language, or low-resource jurisdictions.

\textbf{Label provenance.} The benchmarks derive labels from metadata and
automatic extraction rather than expert annotation produced for evaluation,
which bounds how far absolute numbers should be read as measuring legal
competence.

\textbf{Encoders, not generative models.} We evaluate classification and
first-stage retrieval only. Nothing here speaks to generative legal tasks,
where decoder LLMs dominate and the relevant failure modes differ in kind.

\textbf{Scale.} All models are base-scale.

\section{Future Work}
\label{sec:future}

A subsequent version of this work will address the limitations above directly.
\textbf{Robustness:} five-seed replication following the LexGLUE
protocol~\cite{chalkidis2021lexglue}, with paired bootstrap significance tests
for every model pair. \textbf{Baselines:} LEGAL-BERT, a hierarchical BERT
encoder, and BM25 for retrieval, with parameter budgets matched or reported.
\textbf{Mechanism:} length-stratified evaluation, direct measurement of windows
per document, a throughput-versus-sequence-length curve, and overlap and pooling
ablations. \textbf{Failure analysis:} per-label-frequency breakdowns, SCOTUS
confusion matrices, a retrieval error taxonomy, and qualitative case studies.

Beyond these, \textbf{hybrid architectures} are the most direct test of the
complementarity hypothesis: inserting a small number of attention layers into
an SSM backbone should, if the hypothesis holds, recover performance on
localised signals while retaining linear scaling. \textbf{Adaptive context
selection} would exploit the known sectional structure of legal documents.
\textbf{Multilingual and multi-jurisdictional} evaluation addresses the scope
limitation, and \textbf{retrieval-augmented reasoning} would connect efficient
long-context encoders to generative systems as a scalable first stage.

\section{Conclusion}

We compared selective state-space models with transformer encoders on four
legal classification tasks and two case-retrieval tasks under a shared
pipeline. In these preliminary results, the best SSM is within about one point
of the best transformer on every task, while processing roughly three to four
times as many tokens per second. SSD-Mamba is the stronger SSM throughout,
and DeBERTa remains the stronger choice on SCOTUS. For institutions that must
classify or index legal corpora at scale, this accuracy--efficiency trade-off
is the operative result; establishing \emph{why} it arises, and whether it
holds against legal-domain and hierarchical baselines under full statistical
controls, is the agenda for the next version of this work.

\bibliographystyle{IEEEtran}
\bibliography{references}

\end{document}